**Towards Quantitative Analysis of Deuterium absorption in Ferrite and Austenite during Electrochemical Charging by Comparing Cyclic Voltammetry and Cryogenic Transfer Atom Probe Tomography.**

Dallin J. Barton[1], Dan-Thien Nguyen[1], Daniel E. Perea[2], Kelsey A. Stoerzinger[3], Reyna Morales Lumagui[1], Sten V. Lambeets[1], Mark G. Wirth[2], Arun Devaraj[1*]

[1]Physical and Computational Sciences Directorate, Pacific Northwest National Laboratory, Richland, USA

[2]Environmental Molecular Sciences Laboratory, Pacific Northwest National Laboratory, Richland, USA

[3]School of Chemical, Biological and Environmental Engineering, Oregon State University, Corvallis, OR 97331, United States of America

Corresponding author: arun.devaraj@pnnl.gov

**Abstract**

Hydrogen embrittlement mechanisms of steels have been studied for several decades. Understanding hydrogen diffusion behavior in steels is crucial towards both developing predictive models for hydrogen embrittlement and identifying mitigation strategies. However, because hydrogen has a low atomic mass, it is extremely challenging to detect by most analytical methods. In recent years, cryogenic-transfer atom probe tomography (APT) of electrochemically-deuterium-charged steels has provided invaluable qualitative analysis of nanoscale deuterium traps such as carbides, dislocations, grain boundaries and interfaces between ferrite and cementite. Independently, cyclic voltammetry (CV) has provided valuable analysis of bulk hydrogen diffusion in steels. In this work, we use a combination of CV and cryogenic-transfer APT for quantitative analysis of deuterium pickup in electrolytically charged pure Fe (ferrite) and a model austenitic Fe18Cr14Ni alloy without any second phase or defect trap sites. The high solubility and low diffusivity of hydrogen in austenite versus ferrite are highlighted to result in clear observable signatures in CV and cryogenic-transfer APT results. The remaining challenges and pathway for enabling quantitative analysis of hydrogen pick up in steels is also discussed.

## 1. Introduction

Hydrogen embrittlement (HE) is the reduction in the ductility of an alloy due to elevated internal H concentration. HE occurs in high strength ferritic/martensitic steels and face-centered cubic (FCC) austenitic stainless steels (γ-SS). Fractography with visible-light microscopy shows that γ-SS experiences a ductile-to-brittle transition with increasing H concentration [1, 2]. American Iron and Steel Institute (AISI) standardized 304 stainless steels can achieve over 80 % strain before failure, but cannot achieve a 60 % strain after being charged with H [3-7]. Other stainless steels such as AISI 316, 321, and 347 all suffer a similar degradation in ductility after hydrogen charging [8-11]. This reduction of ductility in γ-SS leads to premature fracture in environments and processing conditions that encourage hydrogen absorption and limits engineering applications. Therefore, understanding the hydrogen distribution in steels is important to develop strategies to mitigate the deleterious effects of hydrogen on the mechanical properties.

Two commonly attributed mechanisms of HE is hydrogen enhanced decohesion (HEDE) and hydrogen enhanced local plasticity (HELP). HEDE suggests that elevated H concentration in materials lowers the cohesive strength causing brittle failure [12, 13]. HELP on the other hand suggests that hydrogen interstitials enhance dislocation mobility, and hence, plasticity [14-16]. In the case of stainless steel, both HEDE and HELP have been identified as the main contributors to HE [17-24]. A predictive model attempting to combine the mechanisms of HEDE and HELP has suggested a strong dependence on both hydrogen concentration and strain [25]. Additional mechanisms such as adsorption-induced dislocation-emission (AIDE) and hydrogen enhancement of the strain-induced generation of vacancies (HESIV) are also postulated to be alternate mechanisms [26, 27]. Any unified theory of HE will require experimental confirmation including the spatial measurement of H distribution in materials.

To verify which hydrogen embrittlement mechanism is responsible for HE of specific steels, hydrogen concentration must be measured with high spatial resolution and compositional sensitivity. Methods such as desorption analysis (thermal [28-34] and electron-induced [35]), and electrochemical methods including cyclic voltammetry (CV) and chronoamperometry [36-39] have been used to measure bulk dissolved hydrogen concentration in alloys. Among these, CV has recently emerged as a valuable ensemble average technique for analyzing hydrogen diffusivity in steels[36-39]. CV involves applying a variable potential difference between the working electrode and reference electrode and measures the resulting current as the potential is varied to study the reduction and oxidation reactions of materials, producing a voltammogram. Features in the voltammogram before and after electrochemical hydrogen charging can be used to analyze hydrogen incorporation in steels [36-39].

Mass spectrometry techniques such as secondary ion mass spectrometry (SIMS) and atom probe tomography (APT) can provide spatially resolved analysis of hydrogen concentration in materials. SIMS has been used to confirm elevated deuterium ($^2H$) concentrations at crack tips of $^2H$-charged γ-SS [40], as well as higher $^2H$ in FCC phases of γ-SS than body-centered cubic (BCC) phases [41, 42]. APT can provide chemical and isotopic information in three-dimensional space at resolutions smaller than 1 nm. Isotope separation allows APT to separate a deliberate charging of $^2H$ from artificial $^1H$ introduced through sample preparation or analysis [43].

In materials with deeper trap sites for hydrogen, observable deuterium can remain even after room-temperature transfer. For example, deuterium in poly-crystalline Si has been unambiguously measured to remain trapped at dislocations and high-angle grain boundaries even after room-temperature vacuum transfer [44]. In the case of Fe, the type of defect can influence the binding energy creating a range of potential traps. Bulk-measured and first-principles calculated hydrogen binding energies of solid solution metal additions usually have a hydrogen binding energy of $-E_b < 30$ kJ/mol [45, 46]. Carbide/Fe [47, 48], oxide/Fe [49], and sulfide/Fe [50] interfaces have stronger hydrogen binding energies, $-E_b > 70$ kJ/mol. Hydrogen binding energies in grain boundaries, voids, and dislocations are roughly found in between these two ranges ($18 < -E_b$ (kJ/mol) $< 78$) [51-53].

Considering the broad range of H binding energies in a material, hydrogen would be expected to persist at high binding energy defects even after hydrogen in lower binding energy sites has diffused away (as occurs during e.g. vacuum transfer followed by APT). Cooling down the sample immediately after hydrogen charging and keeping the sample at cryogenic temperatures during transfer can reduce hydrogen diffusivity, enabling its observation even and lower binding energy sites.

Using the cryogenic plunge freezing process, deuterium has been qualitatively observed in medium and high binding energy traps in various steels. In the last decade, $^2H$ charging-APT studies have been performed with ferritic-martensitic dual-phase steels, precipitation strengthened steels, bainite-austenite and ferrite-austenite dual phase steels [54-66]. These studies, however, cannot yet deliver quantitative information relating specific processing conditions to thermodynamic or kinetic behavior because deuterium is rarely seen at the no or low-energy binding sites. In almost all reports, there is no deuterium found in the matrix of the material after charging, only in higher binding energy locations. To quantitatively characterize specific spatial behavior of traps, a standardized method must be created to compare the relationship between low-energy and high-energy binding sites.

This work combines CV and cryogenic-transfer APT to assess both total and spatially resolved hydrogen uptake. Electropolished needle samples of body centered cubic (BCC) pure Fe and FCC model Fe-18Cr-14Ni (wt. %) alloys were electrochemically charged with deuterium and brought to saturation. Following an accelerated room-temperature vacuum transfer method and a cryogenic freezing, frost removal, and cryogenic transfer method, they were then quantitatively analyzed via APT. This hydrogen quantification from APT was then compared with CV to quantitatively analyze the H-pick up in ferrite and austenite.

**Experimental methods**

Fe-18.0Cr-14.0Ni wt. % (Fe-19.2Cr-13.2Ni at. %, shortened to Fe18Cr14Ni) alloys were induction melted from high-purity elements, then cast and homogenized by remelting five times. The alloys were then cold rolled to a 50% reduction in area and recrystallized to 3 mm thick sheets via annealing at 900 °C for 4 h. Previous crystal structure characterization shows that the austenitized Fe18Cr14Ni was FCC stabilized [67]. Pure Fe rods (99.99 %) were purchased from Goodfellow inc. Bars with a square cross section with an area of 1 $mm^2$ were sectioned from the fabricated pure Fe and Fe18Cr14Ni alloy via electric discharge machining (EDM). The bars were then ground and polished on all sides at decreasing grit size down to 1 µm diamond suspension.

Electrochemical characterizations were conducted using Gamry Potentiostat (Interface 1000). The electrolyte was 0.1 M NaOH (98%, Alfa Aesar) in ultrapure deionized water (Sigma Aldrich). The electrochemical cell consisted of 1 $mm^2$ square cross section bars of Fe-18Cr-14Ni or Fe as working electrode, Ag/AgCl reference electrode (0.281 V *vs.* Standard Hydrogen Electrode (SHE)), and Pt wire as counter electrode. 1 cm length of metal bar was dipped into electrolyte solution. The experimental procedure consisted of 3 stages as follows. First, the electrode was submitted to ten consecutive pretreatment CV cycles between -1.25 V and +0.75 V vs Ag/AgCl at a scan rate of 100 mV $s^{-1}$ (final scan to -1.25 V) to obtain a reproducible sample surface. Next, one cycle of CV between -1.25 V and 0.15 V was recorded before hydrogen charging. This is followed by hydrogen charging by polarizing the sample at a constant voltage of −1.25 V (*vs.* Ag/AgCl), which co-occurs with water reduction on the working electrode. After hydrogen charging, the working electrode is instantly submitted to 5 consecutive CV cycles starting from -1.25 V to 0.15 V (referred to as “after hydrogen charging”). Electrochemical data was processed using Gamry Echem Analyst.

Fabricated bars from the same parent material as measured with CV were then sharpened into needles using a two-step electropolishing technique. The first coarse polishing solution was 25 % perchloric acid in glacial acetic acid. The second fine polishing solution was 2 % perchloric acid in 2-butoxyethanol. A concluding polish was conducted in a Thermo Fisher Hydra plasma FIB with Xe plasma. The voltage for milling was 30 kV at decreasing currents with final milling at 2 kV. APT was performed using a local electrode atom probe (LEAP) 4000XHR (Cameca). All runs were conducted in voltage pulsing mode, with voltage pulse fraction at 20 %. The temperature of every sample during data collection was 40 K. Data were reconstructed using Interactive Visualization and Analysis Software (IVAS 3.8, Cameca).

## 2. Results and Discussion

### *3.1 Cyclic voltammetry of Hydrogen Charged Fe and Fe18Cr14Ni*

Pure Fe and Fe18Cr14Ni (1 $mm^2$ square cross section rods) were electrochemically charged at -1.25 V versus Ag/AgCl for 5 minutes in 0.1 M NaOH. Cyclic voltammograms of Fe and Fe18Cr14Ni electrodes before and after the charging steps are shown in Figure 1(a-b). For Pure Fe, the CV before hydrogen charging is shown in black in Figure 1(a) along with the CV after 5 min hydrogen charging ($1^{st}$ and $5^{th}$ cycles) shown in different shades of red. For Fe18Cr14Ni, the CV before charging and after 5 min hydrogen charging are shown in black and green ($1^{st}$ and $5^{th}$ cycles after charging shown different shades of green) respectively in Figure 1(b). The current density *vs.* time during the 5 min charging of pure Fe and Fe18Cr14Ni are shown in Figure 1(c). The cyclic voltammograms of Fe and Fe18Cr14Ni before hydrogen charging shows a small shoulder at -0.8 V in cathodic scan, which is probably due to the reduction of γ-$Fe_2O_3$ and/or α-FeOOH to $Fe_3O_4$ (Equation 1-2). A broad cathodic peak (c1, c1’) at -1 V to -1.2 V is attributed to the reduction of $Fe_3O_4$ to $Fe(OH)_2$ and/or FeO (Equation 3-4) and possibly to $Fe^0$ (Equation 5-6). The anodic peak (a1, a1’) is attributed to the oxidation of $Fe(OH)_2$ /FeO to $Fe_3O_4$ (Equation 3-4) [68]. After 5 min hydrogen charging of pure Fe, a new anodic peak appears at -0.85 V (marked as a2 in Figure 1(a)), which is assigned to H-oxidation [36, 37]. It should be noted that the signal at -0.85 V also appears in the CV before hydrogen charging but only at much lower intensity. In addition, a new anodic peak (a3) and a high anodic current above -0.4 V is also noticeable in the CV curve from hydrogen charged pure Fe given in Figure 1(a). Both a3 (and a3’) and anodic current above -0.6 V declined with repeated CVs scans. Therefore, we hypothesized that this observation is also related to the oxidation of hydrogen. In contrast, in the hydrogen charged Fe18Cr14Ni, no distinct peak emerged near -0.85 V but overall background current density increased near the hydrogen oxidation peak location as evident in Figure 1(b). Furthermore, the anodic peak a3’ current density increased to a higher value in comparison to a1’ and shifted to higher potential. The higher anodic current is also observed above -0.6 V in anodic scan, like that observed in Fe electrode. With repeated CV scans, the anodic current between -0.8 V to -0.9 V, corresponding to hydrogen oxidation, decreased. Both a3' and anodic current above -0.6 V also declined and is similar to the CV curve before hydrogen charging. This trend is consistent with behaviors observed in the Fe electrode.

$$2Fe_3O_4 + 2OH^- \leftrightarrows 3\ \gamma\text{-}Fe_2O_3 + H_2O + 2e^- \ (1)$$

$$Fe_3O_4 + OH^- + H_2O \leftrightarrows 3\alpha\text{-}FeOOH + e^- \ (2)$$

$$3Fe(OH)_2 + 2OH^- \leftrightarrows Fe_3O_4 + 4H_2O + 2e^- \ (3)$$

$$FeO + 2OH^- \leftrightarrows Fe_3O_4 + H_2O + 2e^- \ (4)$$

$$Fe + 3H_2O \leftrightarrows FeO + 2H_3O^+ + 2e^- \ (5)$$

$$Fe + 4H_2O \leftrightarrows Fe(OH)_2 + 2H_3O^+ + 2e^- \ (6)$$

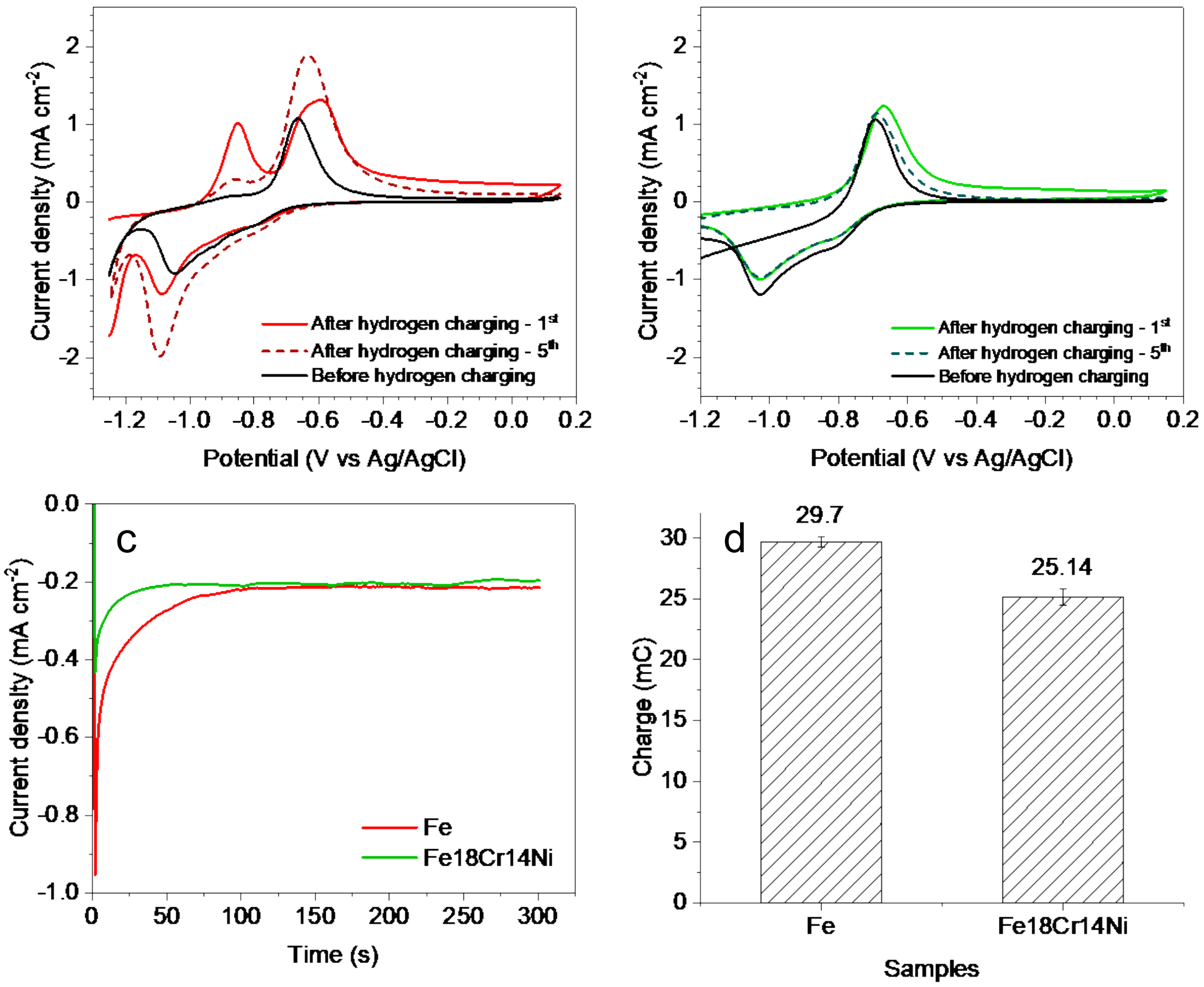

**Figure 1:** Cyclic voltammetry of **(a)** pure Fe and **(b)** Fe18Cr14Ni before and after 5 minutes hydrogen charging at -1.25 V *vs.* Ag/AgCl ($1^{st}$ and $5^{th}$ cycle). **(c)** Chronoamperometric curves for 5 minutes charging and **(d)** charge calculated from curves in **(c)**.

The charge calculated from chronoamperometric curves for 5 minutes hydrogen charging (Figure 1(d)) is recorded to evaluate current response and to determine the amount of charge as a function of time. The Fe electrode shows higher current in the first few cycles of the charging process, corresponding to higher amount of charge transfer during hydrogen charging. The differences in the cyclic voltammogram of Fe and Fe18Cr14Ni electrodes are related to the hydrogen traps present and diffusion from the bulk to the surface, and hydrogen oxidation kinetics (related to the surface composition). We hypothesize that the absence of an overall elevation of current density with no apparent hydrogen oxidation peak (a2) in Fe18Cr14Ni is due to hydrogen diffusivity being lower in austenite in comparison to much faster diffusion out of pure Fe. This might lead to the current from hydrogen oxidation occur over a broader voltage range and the present fast sweep rate. We also do not rule out the role of native $Cr_2O_3$ on Fe18Cr14Ni as a hydrogen permeation barrier as reported in literature [69-72]. First principles calculations determined the diffusion coefficient of hydrogen in $Cr_2O_3$ is 5.03 x $10^{-10}$ $cm^2/s$ at 500 °C, which is threefold slower than that in 316L stainless steel[73]. A more detailed understanding of the hydrogen oxidation mechanism in pure Fe and Fe18Cr14Ni requires intensive experiment investigation in interfacial reactions and will be reported in a future work.

In this study, we focused on the difference in hydrogen diffusion kinetics and storage properties between BCC pure Fe and FCC Fe18Cr14Ni, which can lead to more hydrogen being retained within the FCC lattice of Fe18Cr14Ni. To further verify this hypothesis, atom probe analysis was performed to quantitatively analyze the hydrogen pick up in the electrochemically charged Fe18Cr14Ni and pure Fe.

### *3.2 Establishing a baseline for Hydrogen detected in Fe and Fe18Cr14Ni in Atom Probe Tomography before hydrogen charging*

In APT mass-to-charge spectra, hydrogen background peaks may be observed either from hydrogen that may be introduced into materials during APT sample preparation, or desorption of adsorbed hydrogen on the walls and other components of the atom probe analysis chamber made out of stainless steel [74]. To minimize the detected background hydrogen signal, we performed a parametric study of different variables known to influence hydrogen detection including voltage pulse frequency (kHz) and detection rate (event/pulse). Pulse frequency and detection rates were chosen between 50-200 kHz and 0.5-3.0 % respectively. Both a higher detection rate and higher voltage pulse frequency results in lower background hydrogen detection in both pure Fe (Figure 2a) and Fe18Cr14Ni (Figure 2(b)). The lowest recorded concentration of hydrogen was present at a pulse frequency of 200 kHz and a detection rate of 3.0%, which were the highest detection parameters used. No molecular peaks of hydrogen, at 2 Da, 3 Da, or 4 Da, was observed in either material at all combinations of detection rate and voltage pulse frequencies (mass-to-charge spectra shown in Figure 2(c) for pure Fe and Figure 2(d) for Fe18Cr14Ni).

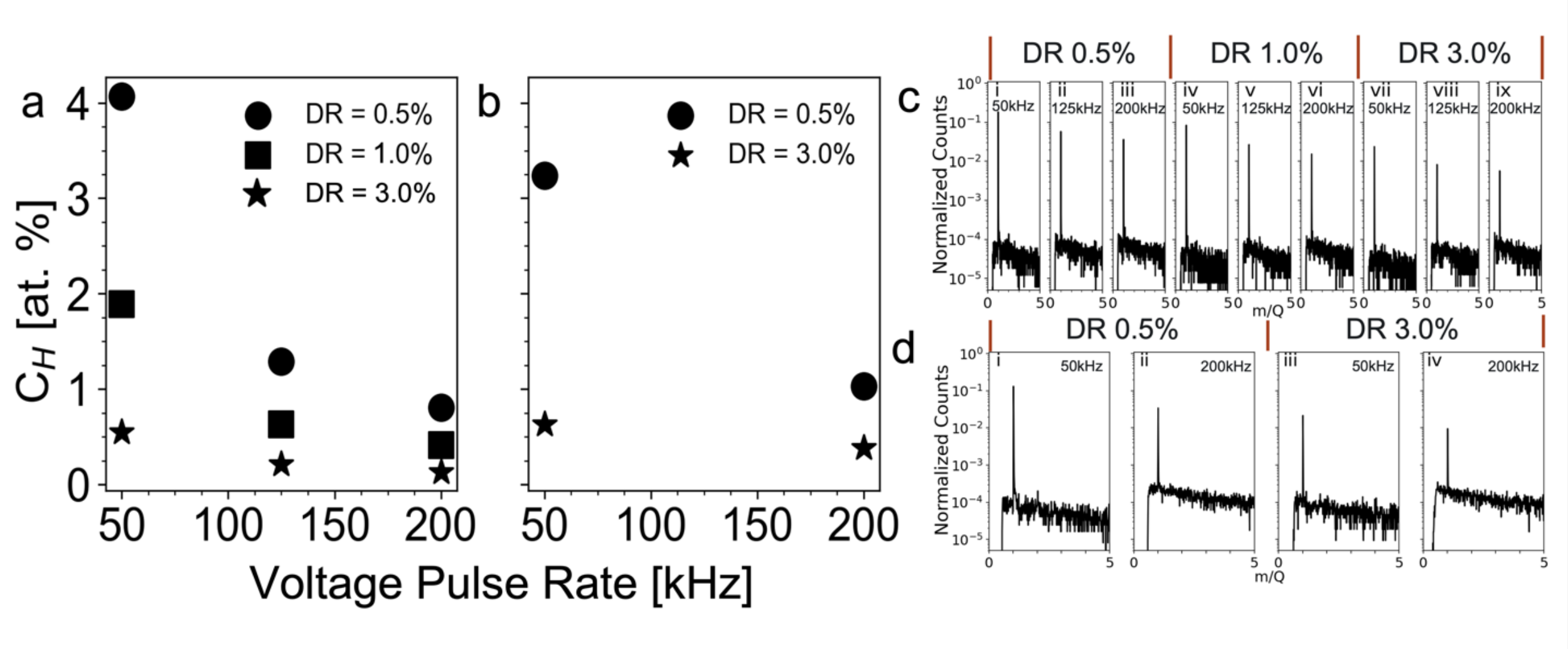

**Figure 2:** APT analysis parameter dependence on hydrogen quantification. Concentration of hydrogen measured by atom probe tomography as a function of voltage pulse frequency (PF) and detection rate (DR) at 20% pulse fraction for **(a)** pure Fe and **(b)** Fe18Cr14Ni. All the data for both materials are collected while analyzing the same needle specimen and parameters were changed during the same APT experiment. The mass-to-charge ratio spectra (m/q) from 0-5 Da given in **(c)** Pure Fe at the parameters of **(ci)** DR 0.5% PF 50 kHz **(cii)** DR 0.5% PF 125 kHz, **(ciii)** DR 0.5% PF 200 kHz, **(civ)** DR 1.0% PF 50 kHz, **(cv)** DR 1.0% PF 125 kHz, **(cvi)** 1.0% PF 200 kHz, **(cvii)** DR 3.0% PF 50 kHz, **(cviii)** DR 3.0% PF 125 kHz, **(cix)** DR 3.0% PF 200 kHz, and **(d)** Fe18Cr14Ni at the parameters of **(di)** 0.5% PF 50 kHz, **(dii)** DR 0.5% PF 200 kHz, **(diii)** DR 3.0% PF 50 kHz, **(div)** DR 3.0% PF 200 kHz. Counts for these mass-to-charge spectra are normalized to the highest count recorded, which is $^{56}Fe^{2+}$.

If the lowest concentration of hydrogen is desired, pulse frequency and detection rate should be maximized; however, this can result in premature fracture of APT needle samples. All Fe18Cr14Ni needles fractured during field evaporation at 3.0 % detection rate and 200 kHz pulse rate before 2 million ions could be collected. While this setting produced the lowest H signal, it was impractical to run at these conditions. A closer inspection of the mass spectra shows that a lower pulse frequency (50 kHz) at a high detection rate (3.0 %) also produced a low hydrogen signal with low background while still resulting in a higher sample yield. Therefore, all subsequent atom probe samples were analyzed at a detection rate and pulse frequency of 3.0 % and 50 kHz to maximize the combination of decreased artificial hydrogen count, lower background, and maintain high sample yield.

### *3.3 Atom Probe Tomography Analysis of Deuterium charged Fe and Fe18Cr14Ni*

Pure Fe and Fe18Cr14Ni, needle samples were analyzed in APT to get the baseline APT results before charging. Then the APT experiments were stopped, and samples were removed to charge $^{2}H$ electrochemically. The samples fixed in a Cu holder were placed in 23 mL of a 0.1 M solution of sodium deuteroxide (NaOD) in deuterium oxide ($D_2O$, 99.9 % pure). The counter electrode wire was Pt. The schematic of the deuterium charging setup is shown in Figure 3(a). A constant direct current bias of -2.2 V was generated by a power supply onto the system for 2 minutes. The steady state (achieved in 10 seconds) reference electrode measurement was -1.6 V vs AgCl for the pure Fe needle and -2.0 V for the Fe18Cr14Ni needle. Here, a -2.2 V is not measured against a working or reference electrode but was the voltage from the power supply on the system. Similar APT works published so far indicate that at this system voltage and time, we can ensure H saturation in the material without producing too many bubbles that could prevent H absorption [54, 65]. In all experiments, <1 bubble per 5 seconds were observed using a visible light stero microscope.

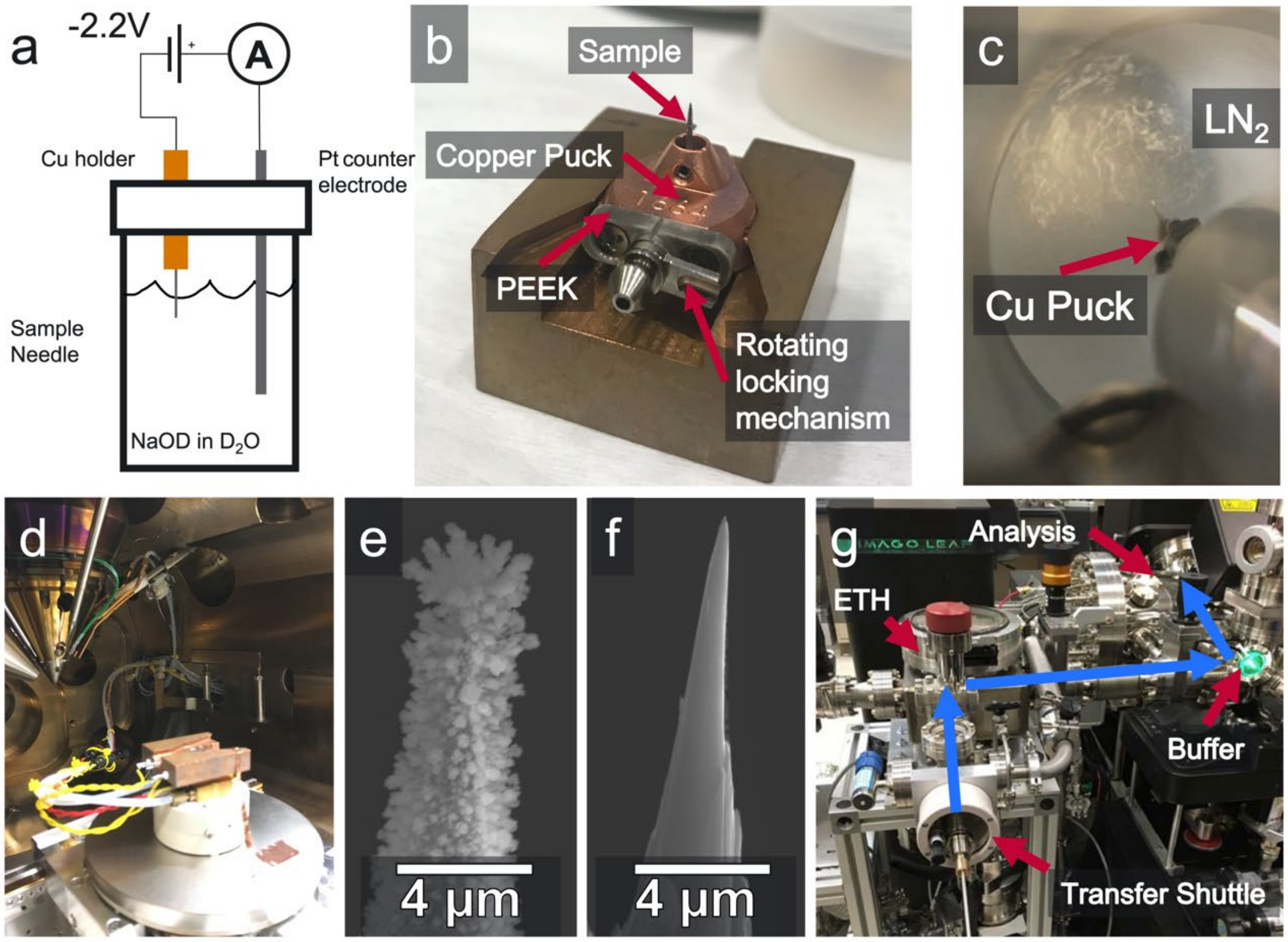

**Figure 3**: Demonstration of $^2$H charging and cryogenic transfer process. **(a)** Charging setup diagram with the sample anode in a Cu holder, Pt wire cathode, and solution. **(b)** Wire attached to the Cu puck modified for thermal isolation against the transfer rod and manipulation by the quorum stage. **(c)** Puck and needle plunging into liquid nitrogen. **(d)** Photograph of cryogenic stage in a FIB/SEM modified to accept APT pucks. **(e)** SEM image of deuterium charged needle sample with frozen condensed water after quenching. **(f)** needle sample after final sharpening. **(g)** Blue arrows indicate path of the sample from the transfer shuttle into the environmental transfer hub (ETH), buffer chamber, and finally, the APT analysis chamber.

After $^2$H charging, the samples were immediately removed from the charging apparatus, dipped three times in a container of pure $D_2O$ to remove any NaOD, blow dried with $N_2$ gas, and placed in an atom probe puck, Figure 3(b). The puck with the deuterium-charged wire sample was then either transferred using a vacuum shuttle at room temperature or at cryogenic temperatures into the atom probe analysis chamber. The time from removing the needle from electrochemical charging solution to plunge freezing was less than 180 seconds.

To achieve cryogenic transfer, the deuterium-charged sample was first plunged into liquid nitrogen (Figure 3(c)). The plunge frozen sample was then transferred into the cryogenic stage of a Thermo Fisher Scientific Quanta FIB/SEM (Figure 3d) using a Quorum shuttle. A minor amount of frost formed during the sample transfer process (Figure 3(e)) which was removed by using 2kV Ga ion beam milling to get a frost-free needle sample (Figure 3(f)). The final needle sample was then transferred into the APT using the Quorum shuttle device, through the cryogenic stage on the environmental transfer hub (ETH), which was maintained at high vacuum and temperature below -150 ˚C. The sample is then transferred from the ETH onto a thermal insulative puck holder on the buffer chamber carousel and then transferred into the APT analysis chamber. This transfer workflow allows the sample to remain at cryogenic temperatures while also maintaining high vacuum, preventing any additional frost formation during transfer process.

A comparison of the mass-to-charge spectra peaks at 0-5 Da from pure Fe before charging (Figure 4(a)) and after 2-minute $^{2}H$ charging and room temperature transfer (indicated by the sun schematic) (Figure 4(b)) show a small but distinct new peak emerging at 2 Da. The composition profile plotted along the z-axis of the APT reconstruction of the 2 min charged-room temperature transferred pure Fe sample is given in Figure 4(c) where 1H is <0.1 at. % and $^{2}H$ concentration is <0.05 at. %.  On the other hand, the comparison of APT mass-to-charge spectra peaks of pure Fe before (Figure 4(d)) and after charging followed by cryogenic transfer (highlighted by a blue snowflake schematic) (Figure 4(e)), show a significant increase in the intensity of the 2 Da peak. The corresponding composition profile from the deuterium charged and cryogenic transferred sample given in Figure 4(f) highlights the increased concentration of $^{1}H$ and $^{2}H$ to 0.2 at. %. The distribution of Fe, $^{1}H$ and $^{2}H$ within the cryogenically transferred deuterium charged pure Fe is shown in Figure 4(g) where a minor tendency for $^{1}H$ and $^{2}H$ to segregate to the poles in the reconstruction is evident. These results highlight the value of cryogenic transfer in retaining more $^{2}H$ within the BCC structured pure Fe APT sample.

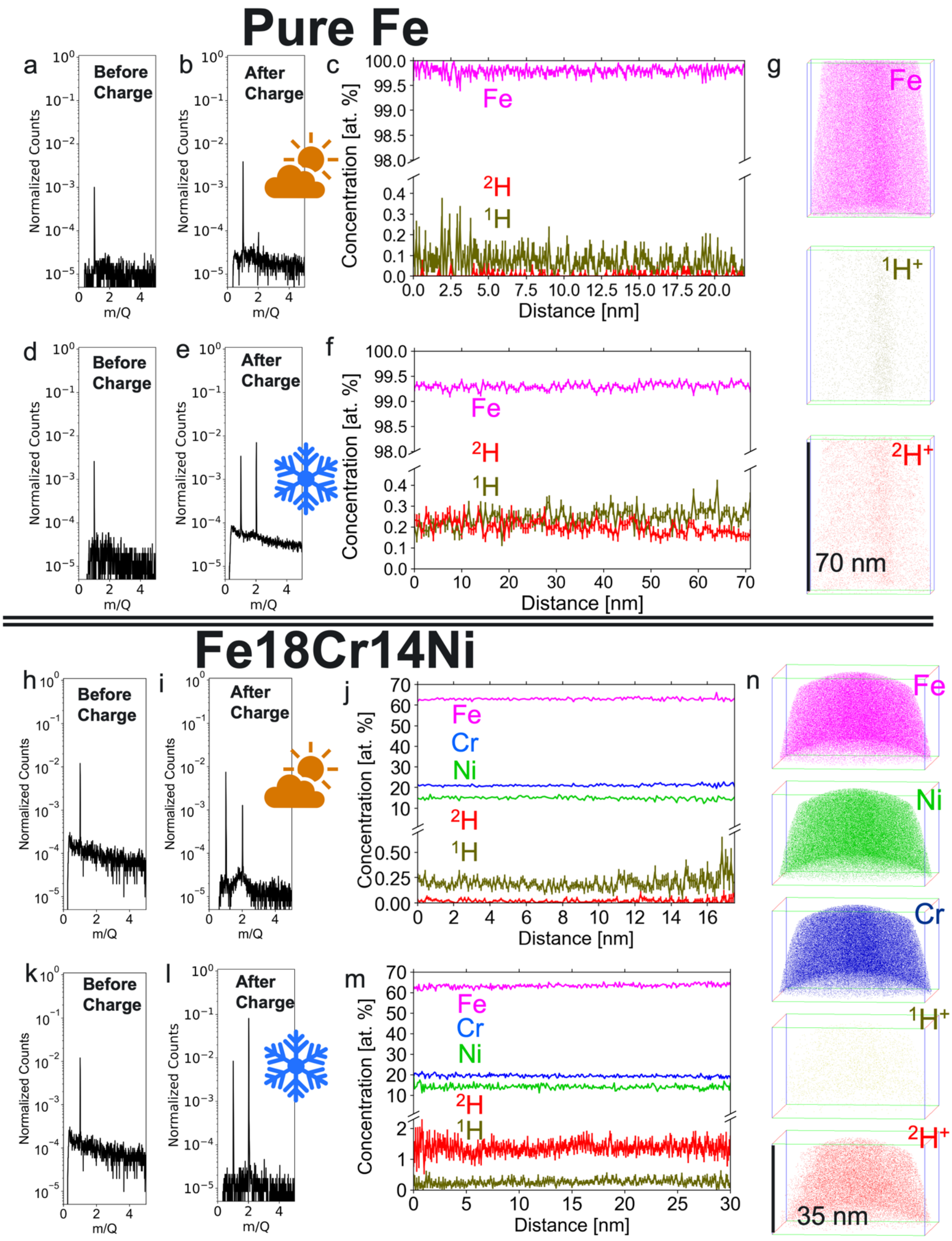

**Figure 4:** Atom probe tomography results of pure Fe and Fe18Cr14Ni charged in 0.1M NaOD in $D_2O$ at -2.2 V for 2 min followed by either room temperature transfer or cryogenic transfer into the APT. APT mass-to-charge ratio spectrum between 0-5 Da for pure Fe **(a)** before charging and **(b)** after 2 min $^2$H charging and room temperature transfer **(c)** composition profile along the z axis of the reconstruction for deuterium charged and room temperature transferred pure Fe. APT mass-to-charge ratio spectrum between 0-5 Da for pure Fe **(d)** before charging and **(e)** after 2 min $^2$H charging and cryogenic transfer **(f)** composition profile along the z axis of the reconstruction for deuterium charged and cryogenic transferred pure Fe. **(g)** distribution of Fe, $^1$H and $^2$H within the APT reconstruction in the deuterium charged and cryogenic transferred pure Fe. APT mass-to-charge ratio spectrum between 0-5 Da for Fe18Cr14Ni **(h)** before charging and **(i)** after 2 min $^2$H charging and room temperature transfer **(j)** composition profile along the z axis of the reconstruction for deuterium charged and room temperature transferred Fe18Cr14Ni. APT mass-to-charge ratio spectrum between 0-5 Da for Fe18Cr14Ni **(k)** before charging and **(l)** after 2 min $^2$H charging and cryogenic transfer **(m)** composition profile along the z axis of the reconstruction for deuterium charged and cryogenic transferred pure Fe. **(n)** distribution of Fe, Cr, Ni, $^1$H and $^2$H within the APT reconstruction in the deuterium charged and cryogenic transferred Fe18Cr14Ni.

Thereafter, we conducted a similar comparison of the effectiveness of room temperature transfer versus cryogenic transfer for Fe18Cr14Ni samples. The comparison of the mass-to-charge spectra peaks at 0-5 Da from Fe18Cr14Ni before charging (Figure 4(h)) and after 2 min deuterium charging and room temperature transfer (Figure 4(i)) show a distinct new peak emerging at 2 Da. The composition profile plotted along the z-axis of the APT reconstruction of the 2 min charged-room temperature transferred Fe18Cr14Ni sample is given in Figure 4(j) where $^1$H concentration is <0.25 at. % and $^2$H concentration is <0.05 at. %. The comparison of 0-5 Da APT mass-to-charge spectra peaks of Fe18Cr14Ni before (Figure 4(k)) and after charging followed by cryogenic transfer (Figure 4(l)), show an even higher intensity for the 2 Da peak. The corresponding composition profile from the deuterium charged and cryogenic transferred sample given in Figure 4(m) show the $^1$H concentration remained close to 0.25 at. % but the $^2$H concentration increased to approximately 1.5 at. % highlighting a significantly high $^2$H pickup and retention within the cryogenically transferred sample. The distribution of Fe, Cr, Ni, $^1$H, and $^2$H within the cryogenically transferred deuterium charged Fe18Cr14Ni is shown in Figure 4(n) where a rather uniform distribution of $^2$H in the matrix was observed. Here again in Fe18Cr14Ni cryogenic transfer process after $^2$H charging led to retention of more $^2$H within the Fe18Cr14Ni versus room temperature transfer. One additional observation is much higher amount of $^2$H concentration within austentic Fe18Cr14Ni versus pure Fe charged for the same time. This difference in observed $^2$H concentration is presumed to be arising from the differences in the solubility and diffusivity of hydrogen within the BCC Fe versus austentic Fe18Cr14Ni.

Gas permeation studies, galvanostatic measurement, thermal desorption spectroscopy, and first principles calculations estimate that the hydrogen diffusion coefficient in ferrite is generally higher than austenite [75-86]. Additionally, the hydrogen solubility in ferrite is lower than in austenite [87]. Therefore, more hydrogen can be diffused into γ-SS than pure Fe during electrochemical charging. Once hydrogen enters the lattice because of electrochemical charging, the subsequent outward diffusion of hydrogen will be much faster from pure Fe than γ-SS. The differences in H solubility and diffusion rate are demonstrated in the CV results of γ-SS compared to pure Fe. The higher diffusion of hydrogen out of the pure Fe leads to a CV peak for stripping hydrogen absorbed onto the steel(a2) whereas no distinct hydrogen stripping feature is present in the austenitic Fe18Cr14Ni. The lower diffusion and higher solubility of hydrogen in austenitic Fe18Cr14Ni allows for the increased current density without a distinct peak at a2. Additionally, this also explained the detection of higher concentration of $^{2}H$ using APT in Fe18Cr14Ni versus pure Fe, charged for the same time and transferred cryogenically into the atom probe analysis chamber.

As a result of this advancement in cryotransfer-APT for quantitative detection of deuterium in the matrix of the alloys, we can develop future systematic studies to quantitatively understand deuterium behavior in materials. Electrochemistry, including solution chemistry, potential, time, and temperature may be varied. Differences between trap sites may be measured with greater rigor. Diffusion time constants may be assessed by varying times to quench after charging. Spatial diffusion may be measured as a function of distance enabled by milling the tip via cryo-FIB.

## 3. Conclusions:

In summary, using CV and cryogenic-transfer APT, we approach quantitative analysis of the hydrogen pickup in body-centered cubic structured pure Fe and face centered cubic austenitic Fe18Cr14Ni model alloy. CV results showed clear differences in hydrogen oxidation from hydrogen charged pure Fe being faster than from Fe18Cr14Ni alloy. Using deuterium as a tracer, cryogenic transfer versus room temperature transfer to APT after deuterium charging, always resulted in detecting a higher concentration of $^{2}H$ in both pure Fe and Fe18Cr14Ni alloy. The overall higher concentration of $^{2}H$ detected after charging and cryogenic transfer to APT in Fe18Cr14Ni was attributed to the lower hydrogen diffusion coefficient and higher solubility of hydrogen in austenite than ferrite and potentially the native chromium oxide on surface acting as a diffusion barrier for hydrogen. We believe correlating CV and cryogenic transfer atom probe tomography can be highly informative in understanding the diffusion and trapping of hydrogen in Fe-based alloys with and without defects and trapping sites.

## 4. Acknowledgements

This research was supported by the U.S. Department of Energy, Office of Science, Basic Energy Sciences, Materials Sciences and Engineering Division as a part of the Early Career Research program (FWP # 76052). This work was also supported in part by the U.S. Department of Energy, Office of Science, Office of Workforce Development for Teachers and Scientists (WDTS) under the Science Undergraduate Laboratory Internships Program (SULI). KAS was supported by the U.S. Department of Energy, Office of Science, Basic Energy Sciences, Chemical Sciences, Geosciences, & Biosciences Division as a part of the Early Career Research program (DE-SC0022970). The APT was conducted using facilities at Environmental Molecular Sciences Laboratory (EMSL), which is a DOE national user facility funded by Biological and Environmental Research Program located at Pacific Northwest National Laboratory. The authors would like to acknowledge Jack Grimm for his assistance in charging and cryogenic transfer and the EMSL machine shop for sample fabrication.